\documentclass[pra,aps,showpacs,floatfix,twocolumn]{revtex4-2}
\usepackage{amsmath}
\usepackage{amsfonts}
\usepackage{amssymb}
\usepackage{revsymb}
\usepackage{graphicx}
\usepackage{physics}
\usepackage{xcolor}
\usepackage{booktabs}
\usepackage{tabularx}
\usepackage{url}

\newcommand{\sref}[1]{Sec. \ref{#1}}
\newcommand{\eref}[1]{Eq.~(\ref{#1})}
\newcommand{\fref}[1]{Fig.~\ref{#1}}
\newcommand{\tref}[1]{Table~\ref{#1}}
\newcommand{\Lu}{$^{176}\mathrm{Lu}^+$}
\newcommand{\fresA}{$10\,491\,519\,945.228\,82(38)\,$}
\newcommand{\fresB}{$11\,290\,004\,289.881\,61(36)\,$}

\begin{document}

\title{Precision measurement of the \textsuperscript{176}Lu\textsuperscript{+}  \textsuperscript{3}D\textsubscript{1} microwave clock transitions}

\author{M. D. K. Lee}
\affiliation{Centre for Quantum Technologies, National University of Singapore, 3 Science Drive 2, 117543 Singapore}
\author{Qi Zhao}
\affiliation{Centre for Quantum Technologies, National University of Singapore, 3 Science Drive 2, 117543 Singapore}
\author{Qin Qichen}
\affiliation{Centre for Quantum Technologies, National University of Singapore, 3 Science Drive 2, 117543 Singapore}
\author{Zhao Zhang}
\affiliation{Centre for Quantum Technologies, National University of Singapore, 3 Science Drive 2, 117543 Singapore}
\author{N. Jayjong}
\affiliation{Centre for Quantum Technologies, National University of Singapore, 3 Science Drive 2, 117543 Singapore}
\author{K. J. Arnold}
\affiliation{Centre for Quantum Technologies, National University of Singapore, 3 Science Drive 2, 117543 Singapore}
\affiliation{Temasek Laboratories, National University of Singapore, 5A Engineering Drive 1, Singapore 117411, Singapore}
\author{M. D. Barrett}
\affiliation{Centre for Quantum Technologies, National University of Singapore, 3 Science Drive 2, 117543 Singapore}
\affiliation{Department of Physics, National University of Singapore, 2 Science Drive 3, 117551 Singapore}

\begin{abstract}
We report precision measurement of the unperturbed ${^{3}}D_1$ microwave transition frequencies in $^{176}\mathrm{Lu}^+$ to a fractional uncertainty of $4\times10^{-14}$. We find the $F = 8$ to $F = 7$ hyperfine transition frequency to be \fresA Hz and the $F = 7$ to $F = 6$ transition frequency to be \fresB Hz. At this precision we are able to observe the hyperfine-mediated effects in the ratio of the quadrupole shifts, from which we can directly infer the residual quadrupole moment after $^3D_1$ hyperfine averaging. We find a residual quadrupole moment of  ${-2.48(23)\times10^{-4}}\,e a_0^2$, consistent with a previous assessment using a different and less direct method.  With the unperturbed microwave frequencies accurately known, the residual quadrupole shift for a $^{176}\mathrm{Lu}^+$ ($^3D_1$) optical frequency standard can henceforth be readily evaluated to $<10^{-20}$ uncertainty by routine ${^{3}}{D}_1$ microwave spectroscopy. 
\end{abstract}

\maketitle

\section{Introduction}
Optical and microwave frequency standards utilizing states with angular momentum $J>1/2$ are affected by quadrupole shifts arising from the coupling between each states' electronic quadrupole moment and electric field gradients. These shifts can be a significant obstacle in ion-based clocks~\cite{diddams2001optical,dube2005electric}, but are typically suppressed by averaging over magnetic field magnetic field orientations~\cite{itano2000external}, over all magnetic substates~\cite{dube2005electric}, dynamic-decoupling~\cite{shaniv2019quadrupole},  or over hyperfine states $\ket{F,m=0}$ for all $F$~\cite{barrett2015developing}. 

The $^{176}\mathrm{Lu}^{+}$ $^1S_0$ to $^3D_1$ transition has several properties desirable for an optical frequency reference, which include a long excited state lifetime, low blackbody radiation shift~\cite{arnold2018blackbody}, magnetic field insensitive ($m=0$) clock states, and a large atomic mass which reduces sensitivity to motional effects. While the quadratic Zeeman and quadrupole shifts of the $^3D_1$ state initially pose a challenge, hyperfine averaging (HA) mitigates these effects by realizing an effective $J=0$ state~\cite{barrett2015developing}. HA in $^{176}\mathrm{Lu}^{+}$ is realized either as an average of optical frequencies or, more practically, by a hybrid microwave and optical interrogation technique first demonstrated in~\cite{kaewuam2020}. Although HA practically eliminates the leading quadrupole shift, a residual quadrupole moment (RQM), which arises from hyperfine-mediated mixing between fine-structure manifolds~\cite{zhiqiang2020hyperfine,beloy2017hyperfine}, remains. The quadrupole shift due to the RQM on the optical frequency is typically small, $<10^{-18}$ fractionally, but must be assessed at the $10^{-19}$ level of systematic uncertainty now being reached~\cite{zhiqiang2024}. The RQM for a $^{176}\mathrm{Lu}^+ ({^3}D_1$) frequency standard has been estimated to be $-2.54(25)\times10^{-4}\,e a_0^2$ ~\cite{zhiqiang2020hyperfine}, where $e$ is the elementary electron charge and $a_0$ is the Bohr radius, but an effective means of measuring the electric field gradient {\it in situ} has been lacking. 

The $^{176}\mathrm{Lu}^{+}$ ${^3}{D}_1$ microwave clock transition frequencies $\ket{F,m_F}=\ket{7,0}$ to $\ket{8,0}$ ($\ket{6,0}$), denoted $f_1$ ($f_2$) respectively,  are routinely tracked with $\sim$mHz accuracy as part of $^1S_0$ to $^3{D}_1$ optical clock operation. The leading systematic shifts influencing these microwave transition frequencies are the quadratic Zeeman shift and the quadrupole shift. The quadratic Zeeman shift is readily evaluated using the known quadratic Zeeman coefficients and the magnetic field amplitude inferred from a measured Zeeman splitting of $m=\pm1$ magnetic field sensitive states. With accurate knowledge of the unperturbed microwave clock transition frequencies, the residual quadrupole shift on the HA optical clock frequency may be readily obtained from the quadrupole shifts on the microwave frequencies~\cite{kaewuam2020quad,zhiqiang2020hyperfine}.

Here we report precision measurement of the unperturbed $^3{D}_1$ microwave clock transition frequencies. First, the quadratic Zeeman coefficients are found by microwave Ramsey spectroscopy over a range of magnetic field amplitudes from 0.13 mT to 2.6 mT. The resulting quadratic Zeeman coefficients are consistent with previously reported values~\cite{zhiqiang2024}, but with one order of magnitude reduction in the uncertainty. We present a scheme to precisely set the orientation of the magnetic field using spectroscopic measurements on the ion. The orientation of the electric field gradient is characterized by measuring quadrupole shifts over a range of applied magnetic field directions. Finally, the unperturbed microwave transition frequencies are measured by averaging the quadrupole shift over three orthogonal directions of the magnetic field. 

Applying the newly measured microwave frequencies, we demonstrate that for a large applied electric field gradient, the quadrupole shift can be evaluated with sufficient precision to resolve hyperfine-mediated effects. From the ratio of quadrupole shifts on the two microwave transitions, the RQM remaining after HA can be inferred, providing an independent experimental validation of the previously reported value derived from a combination of $g$-factor measurements and calculated matrix elements~\cite{zhiqiang2024}. 

\section{Experimental system}

Experiments are performed with a single $^{176}\mathrm{Lu}^+$ ion in the apparatus denoted `Lu-1' in previous work~\cite{EMM}.   The ion trap consists of two axial end caps separated by $\sim 2.7\,$mm and four rods arranged on a square with sides of 1.2 mm length~\cite{zhiqiang2020hyperfine,zhao2025lande}. The rf potential at frequency $\Omega_\mathrm{rf}=2\pi \times 9.4\,$MHz is delivered via a quarter-wave helical resonator to two diagonally opposite rods. Under typical operating conditions, static potentials of 9 V are applied on the end caps and 0.5 V on the two other diagonally opposite rods, which, together with the rf-potential, results in measured trap frequencies for a single \Lu ion of $2\pi \times (648,614,177)$ kHz for the two radial and axial modes respectively.

The atomic-level structure of $^{176}\mathrm{Lu}^+$ is shown in \fref{fig:schematic}(a). Repump lasers at 350, 622, and 895 nm initialize population in the $^3D_1$ state. A laser at 646 nm provides Doppler cooling and state detection for the $^3D_1$ state by fluorescence collected onto a single photon counting module. An additional $\pi$-polarized 646-nm beam addressing $F = 7$ to $F' = 7$ facilitates state preparation into $\ket{^3D_1,7,0}$. Lasers at 848 nm and 804 nm drive the $^1S_0$ to ${^3}D_1$ and $^1S_0$ to ${^3}D_2$ clock transitions, respectively. Two microwave horn antennas external to the vacuum chamber are used to drive the $\Delta m=0,\pm 1$ microwave transitions between hyperfine levels. The microwave polarization at the ion is coarsely controlled by manual repositioning of the horns as required. Three orthogonal pairs of coils produce a magnetic field up to 0.2 mT at the ion, which can be oriented in any direction.

With the exception of auxiliary experiments to determine the magnetic field alignment, the primary measurements consist of Ramsey spectroscopy of the ${^{3}}D_1$ microwave clock transitions. For all measurements the interrogation sequence begins only after conditional detection of the ion in the ${^3}D_1$ state by fluorescence detection with repump and cooling beams on. The ion is then Doppler cooled for 5 ms and optically pumped for 1 ms to the $\ket{^3D_1,7,0}$ state. Either the $\ket{7,0}$ to $\ket{8,0}$ or $\ket{7,0}$ to $\ket{6,0}$ microwave clock transition is interrogated by Ramsey spectroscopy with a typical $\frac{\pi}{2}$ pulse time of 5 ms and Ramsey time of $T=5\,$s if at low (0.1 mT) magnetic field. State detection of $^3D_1$ is performed after shelving $\ket{7,0}$ population to ${^1}S_0$ on the 848 nm optical clock transition.

A frequency discriminator is derived by alternately stepping the microwave frequency $\pm \frac{1}{4T}$, corresponding to one quarter of a Ramsey fringe of $\frac{1}{T}$ and taking the difference. After averaging $N$ cycles, the microwave frequency is updated to track the center Ramsey fringe. We interleave additional microwave Rabi spectroscopy measurements to track the $\ket{^3D_1,6,\pm 1}$ Zeeman splitting, which requires only a small fraction of the overall duty cycle. The magnetic field amplitude is inferred from the Zeeman splitting using the previously reported Land\'e g-factor~\cite{zhiqiang2020hyperfine}.

\begin{figure}[th]
\begin{center}
\includegraphics[width=1.0\linewidth]{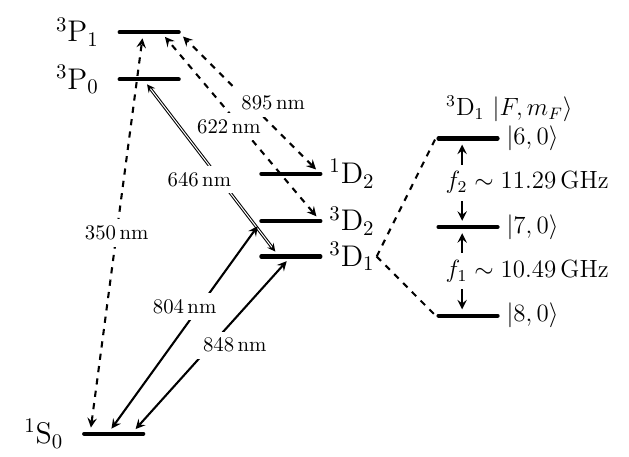}
\caption{\label{fig:levels} Atomic-level structure of $^{176}$Lu$^+$ showing the wavelengths of repump (dashed lines), cooling/detection (double line), and clock (solid lines) transitions used.  Microwave clock transitions $f_1$ and $f_2$ measured in this work}
\end{center}
\end{figure}

\section{Theory \label{sec:theory}}
The quadrupole shift arises from coupling between the electric field gradient and electronic quadrupole moment of the atom~\cite{itano2000external}.  In the  principal-axis (primed) frame, the static electric potential in the neighborhood of the trapped ion has the simple form
\begin{equation}
\Phi(x^\prime, y^\prime, z^\prime) = A [(x^{\prime 2} + y^{\prime 2} +z^{\prime 2} ) + \epsilon (x^{\prime 2} - y^{\prime 2} )]. \label{eq:potential}
\end{equation} 
We stress that the principal axes are determined by the orientation of the static quadrupole field.  They are distinct from the principal axes of the trap, which includes the pseudo-potential confinement and defines the eigenmodes of motion.  These need not be aligned to each other as explicitly illustrated in \sref{sec:results}.
The quadrupole shift in the laboratory frame where the $z$ axis is aligned to the magnetic field is given by~\cite{itano2000external}
\begin{multline}
h f_Q (F, m_F) =  C_{F,m_F} \Theta(J) \\
\quad \times A[(3 \cos^2{\beta} - 1) - \epsilon \sin^2{\beta} (\cos^2 \alpha - \sin^2 \alpha)].\label{eq:quad}
\end{multline} 
Here $\alpha$ and $\beta$ are the Euler angles for rotation of the principal axis frame to the laboratory frame following the conventions of~\cite{itano2000external}, 
\begin{multline}
C_{F,m_F} = (-1)^{2F-m_F+I+J}(2F+1)\begin{Bmatrix} F&2&F\\J&I&J \end{Bmatrix}\\ 
\times\begin{pmatrix} F&2&F\\-m_F&0&m_F \end{pmatrix} \begin{pmatrix} J&2&J\\ -J&0&J \end{pmatrix}^{-1},\label{eq:cfmf}
\end{multline}
and $\Theta(J)$ is the quadrupole moment for the fine-structure level defined by
\begin{equation}
\Theta(J) =  \begin{pmatrix} J&2&J\\ -J&0&J \end{pmatrix} \langle J || \Theta^{(2)}|| J \rangle.
\end{equation}

As the laboratory frame changes with the orientation of the magnetic field, it is to be understood that it is distinct from the space-fixed axes shown in \fref{fig:schematic} used to define the overall geometry of the apparatus including the orientation of the magnetic field settings.

The $^{176}$Lu$^+$ ${^{3}}{D}_1$ quadrupole moment has been previously reported as $\Theta(^3{D}_1) = 0.638\,62(74) \, e a_0^2$ ~\cite{kaewuam2020quad}. The coefficients for the quadrupole shifts on the $f_1$  and $f_2$ microwave transitions are $C_1 \equiv  C_{7,0} - C_{8,0} = \frac{8}{5}$ and $C_2 \equiv C_{6,0} - C_{7,0} = -\frac{7}{5}$, respectively. Neglecting the small rf contribution to the axial confinement, the gradient can be estimated by $A = -\frac{m \omega_a^2}{4 e}$, where $m$ is the mass of the ion and $\omega_a = 2\pi\times 177\,$kHz is axial confinement frequency for our typical applied dc voltages. The corresponding quadrupole shift is $\approx 0.78\,$Hz for the $f_1$ transition at the extremum when the magnetic field is aligned to the ion trap axis ($\beta=0$).  

It is of interest to consider the ratio of the quadrupole shifts on the two microwave transitions, which would be precisely $r= \frac{C_1}{C_2} = -\frac{8}{7}$ in the absence of hyperfine mediated effects. Following the treatment in ~\cite{zhiqiang2020hyperfine}, hyperfine mediated effects lead to a modified ratio
\begin{equation}
\tilde{r} = \frac{C_1 \Theta(^3{D}_1) -\frac{16}{175} \beta_{1,2}^Q}{C_2 \Theta(^3{D}_1) -\frac{3}{25} \beta_{1,2}^Q} \approx -1.146\,26(33) \label{eq:ratio}
\end{equation}
where $\beta_{1,2}^Q = -0.0133(13)\,e a_0^2$ is the dominant RQM correction due to mixing with the ${^3}D_2$ state reported in~\cite{zhiqiang2020hyperfine}. The uncertainty of $\beta_{1,2}^Q$ is estimated to be $10\%$ limited by theory.  Experimental determination of the ratio with $\sim0.03\%$ uncertainty would provide direct experimental validation of the residual quadrupole shift assessment which has been determined from a combination of $g$-factor measurements and theory~\cite{zhiqiang2020hyperfine}. 

\section{Calibration of Magnetic Field Alignment\label{sec:align}}

\begin{figure}[t]
\begin{center}
\includegraphics[width=1.0\linewidth]{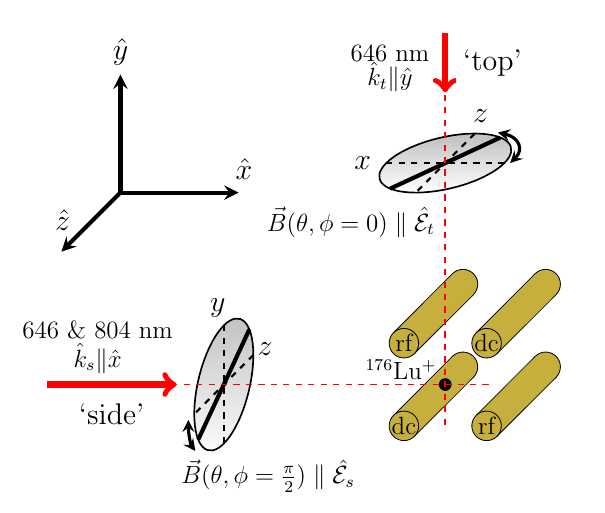}
\caption{\label{fig:schematic} Schematic of the ion trap geometry and laser orientations for the beams critical to the alignment procedures discussed in \sref{sec:align}.  Polarizers, depicted by the grey ellipses, lie in the coordinate planes of the space-fixed axes with solid black lines depicting possible settings of the polarizers and magnetic fields as discussed in the text.  As indicated, the rods of the ion trap lie along the $z$-direction and the laser beams propagate along the $x$ and $y$ directions. The magnetic field orientation described by the spherical coordinates $(\theta,\phi)$, which are polar and azimuthal angles defined in the usual the way with respect to the space-fixed Cartesian coordinates shown. }
\end{center}
\end{figure}

The quadrupole shift averaged over any three mutually orthogonal magnetic field orientations is zero~\cite{itano2000external}. An essential part of this work is to precisely calibrate the angular orientation of the magnetic field, given by the spherical coordinates $(\theta,\phi)$ in the space-fixed frame $(x,y,z)$, as illustrated in \fref{fig:schematic}. The key technique is to use measurements on the ion to always align the magnetic field to the linear polarization of a nominally $\pi$-polarized 646-nm laser used for optical pumping (OP) which provides a precise reference in the space-fixed frame. As shown in \fref{fig:schematic}, the wave vector $\hat{k}_t$ of the OP laser from the 'top' defines the $\hat{y}$ axis and by rotating the angle of the `top' linear polarizer, the magnetic field $\vec{B}$ can be set to any angle $\theta$ in the $xz$-plane $(\phi=0)$ by aligning to the laser polarization $\hat{\mathcal{E}}_t$. By sending the OP laser from the side, $\hat{k_s}=\hat{x}$, we can set $\vec{B}$ to any angle $\theta$ in the $yz$-plane $(\phi=\frac{\pi}{2})$ referenced by the rotation angle of the `side' polarizer. We ensure that $\hat{k}_s\parallel \hat{x}$ by auxiliary measurements on the ion making use of the selection rules for the electric quadrupole (E2) transition at 804-nm. 

\begin{figure}[t]
\begin{center}
\includegraphics[width=1.0\linewidth]{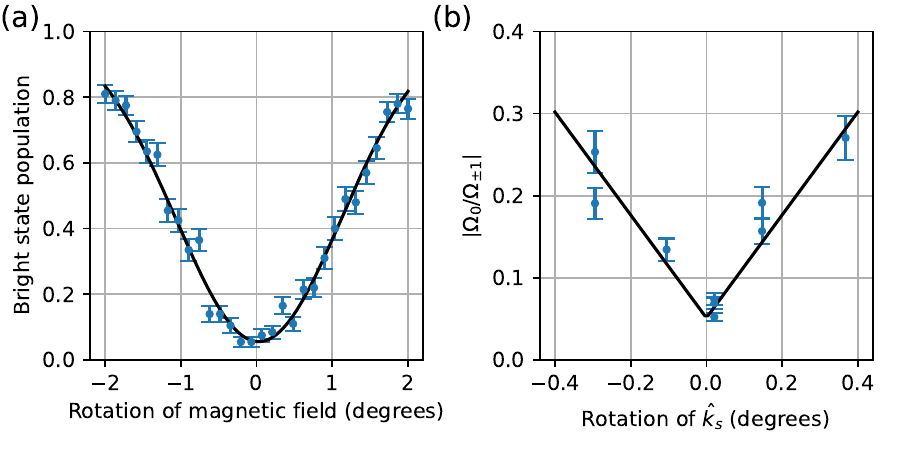}
\caption{\label{fig:alignment} (a) Example of alignment of the magnetic field to the linear OP polarization. (b) Example of alignment of $\hat{k}_{s}$ to the magnetic field via the 804-nm transition E2 coupling.}
\end{center}
\end{figure}

First we describe the procedure to align the magnetic field to the linear polarization of a $\pi$-polarized 646-nm OP laser.  This is done by scanning the angular deflection of the magnetic field and observing the depumping rate out of the $\ket{^3D_1,7,0}$ state from the OP laser due to small angle misalignment of the magnetic field relative to the laser polarization. The depumping rate is measured by first preparing in $\ket{^3D_1,7,0}$, pulsing on only the 646-nm OP beam addressing $F = 7$ to $F' = 7$ for a fixed duration, shelving the remaining $\ket{^3D_1,7,0}$ population to $\ket{^1S_0,7,0}$ with the 848-nm clock laser, and finally state detection of $\ket{^3D_1}$. \fref{fig:alignment}(a) shows the observed population for a typical scan of the magnetic field deflection. The typical fit uncertainty is 0.05$^{\circ}$, and repeated scans yield optimum values consistent with this uncertainty. Small angle scans are performed in both $\theta$ and $\phi$ to optimize alignment of the magnetic field to the 646-nm laser polarization. 

The space-fixed axes $(x,y,z)$ are defined by the OP laser labeled `top' as shown in \fref{fig:schematic}. The optics assembly is rigidly attached to the vacuum chamber with the laser light delivered by optical fiber. The light is collimated and passed through a high extinction ratio ($>10^5$) linear film polarizer in a precision rotation mount before being focused onto the ion by a lens. The rotation mount has a vernier scale by which the rotation angle can be reliably set with 5 arcmin precision. The $\hat{y}$ axis is defined as the propagation direction $\hat{k_t}$, the $\hat{z}$ axis by the polarization of the OP laser when the polarizer is set to a reference angle approximately aligned to the ion trap axis, and the $\hat{x}$ axis as the polarization of the OP laser with the polarizer rotated $90^\circ$ with respect to the $\hat{z}$ setting. By setting the polarizer angle and then aligning the magnetic field to the OP laser polarization, we are then able to precisely set the magnetic field to any direction in the $xz$-plane. We bound systematic uncertainty by rotating the polarizer by 180$^\circ$ and repeating the magnetic field alignment, for which we find a difference of 0.4$^\circ$ in the re-optimized field direction. We take this as the systematic uncertainty in the orthogonality of $\hat{x}$ and $\hat{z}$ magnetic field settings.

To access a third orthogonal direction, we require an additional 646 nm OP laser from another direction, labelled `side' in \fref{fig:schematic}. Here we describe the procedure to ensure the propagation direction of the laser, $\hat{k_s}$, is well aligned in the coordinate frame defined by the `top' beam.  For the `side' optics assembly, both 646 nm and 804 nm light are delivered over the same optical fiber and the entire assembly is on a tip/tilt stage. Again a linear film polarizer in a precision rotation mount determines the polarization $\hat{\mathcal{E}}_s$. We use the relative coupling strengths $\Omega_{\Delta m}$ for $\Delta m = [0,\pm1]$ on the E2 clock transitions at 804 nm to align $\hat{k}_s$ to $\hat{x}$ by minimizing the ratio $\left| \frac{\Omega_{0}}{\Omega_{\pm1}} \right|$ using the tip/tilt stage. With $\vec{B}$ aligned to $\hat{x}$, the 804 nm laser polarization is set to $\hat{y}$ ($\hat{z}$) and the the rotation of $\hat{k}_s$ about $\hat{z}$ ($\hat{y}$) is adjusted to find the minimum ratio, as shown in \fref{fig:alignment}(b) for one of the rotation directions. Next, with $\vec{B}$ aligned to $\hat{z}$ as referenced from the `top', we find the rotation angle of the `side' polarizer that minimizes 646-nm depumping. A 90${^\circ}$ rotation of the polarizer from this position sets the `side' OP laser polarization to $\hat{y}$.  As a final consistency check, we compare the alignment of $\vec{B}$ when set to $\hat{z}$ referenced from either the `top' or `side' OP laser. These two methods agree to within $0.5^\circ$, which we take as systematic uncertainty in the coincidence of the `side' reference frame relative to the `top' reference frame. 

To summarize, by referencing the magnetic field to the OP laser polarization from either the `top' or `side', we are able to precisely set the field direction in the spherical coordinates $(\theta,\phi)$ for any $\theta$ where $\phi$ is restricted to 0 or $\frac{\pi}{2}$, when referenced from the `top'  or `side' ports respectively. The systematic uncertainty in the orthogonality of $\hat{x}$, $\hat{y}$, and $\hat{z}$ magnetic field orientations is estimated to be $0.5^{^\circ}$.

\section{Measurement of quadratic Zeeman coefficients\label{sec:qzc}}

Measured values for quadratic Zeeman coefficients of the ${^{3}}D_1$ microwave transitions, $\alpha_1 $ and $\alpha_2$, were previously reported in the supplemental material of ~\cite{zhiqiang2024}. Here we remeasure these coefficients with improved precision. Measurements of both $f_1$ and $f_2$ are taken at magnetic field amplitudes near 0.13, 0.8 and 2.6 mT with Ramsey interrogation times of 8 s, 1 s, and 250 ms, respectively. Short time scale magnetic field noise of $\sim 10$ nT limits the Ramsey coherence time at larger magnetic fields due to the increased linear sensitivity, $\frac{d f_i}{dB}= 2 \alpha_i B$ for the transitions $i={1,2}$. For magnetic fields $>$0.13 mT, neodymium permanent magnets external to the experiment chamber provide additional bias while the coil pairs provide fine control of the magnetic field orientation and amplitude. For all measurements to determine the quadratic Zeeman coefficients the magnetic field is aligned to the OP laser reference with the polarizer position at a fixed angle throughout, corresponding to $(\theta,\phi) \approx (54^\circ,0)$.

\begin{figure}[t]
\begin{center}
\includegraphics[width=1.0\linewidth]{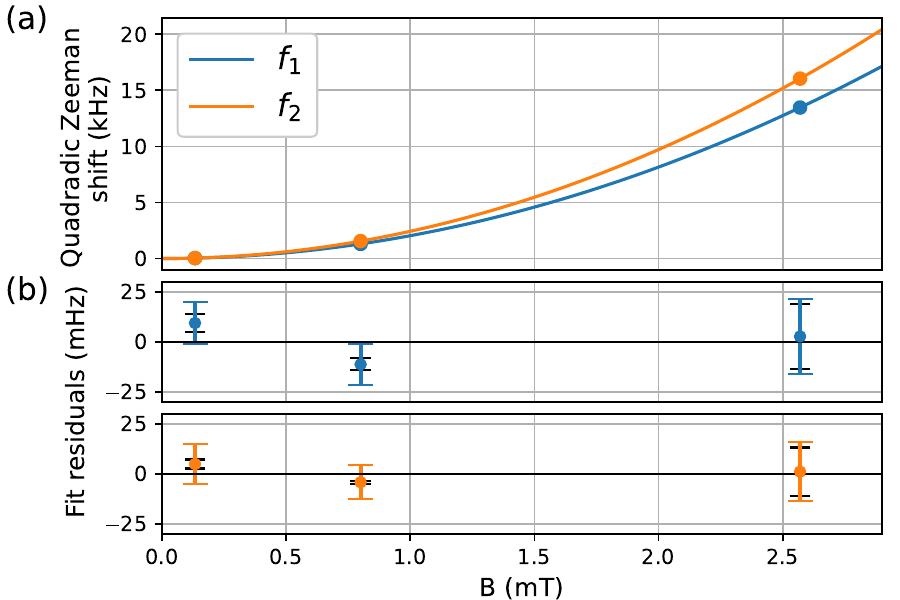}
\caption{\label{fig:qzfit} Evaluation of microwave quadratic Zeeman coefficients. (a) Measurements of both ${^{3}}D_1$ microwave transitions measured as a function of magnetic field amplitude. Solid lines are a quadratic fit. (b) Residuals with respect to the fit. The larger error bars are the combined total uncertainty, and the black error bars are the statistical component only.}
\end{center}
\end{figure}

Results are shown in \fref{fig:qzfit}(a) where each point represents the mean from several hours of averaging. The quadratic Zeeman coefficients are determined from a quadratic fit after subtracting the quartic component, $\lambda_i B^4$, where the coefficients $\lambda_1 = -146\,\mu\mathrm{Hz}/\mathrm{mT}^4$ and $\lambda_2 = -311\,\mu\mathrm{Hz}/\mathrm{mT}^4$ are evaluated from perturbation theory. The quartic component corrects the highest field measurements at the level of 1$\sigma$.  The total and statistical uncertainties for each measurement can be seen in the fit residuals shown in \fref{fig:qzfit}(b). The dominant systematic is the uncertainty in the quadrupole shift when repeating the alignment of the magnetic field between configurations. We expect $\theta\approx54^\circ$ is near to the angle $\beta=1/\sqrt{3}$ where the quadrupole shift, \eref{eq:quad}, vanishes but has linear sensitivity to variation in the angle. We estimate $\pm0.5^\circ$ reproducibility of the field alignment for the high field measurements which contributes 10 mHz uncertainty through the quadrupole shift. Although the  $(\theta,\phi) \approx (54^\circ,0)$ field orientation is standard for clock operation in our lab, we note that, in retrospect, the quadrupole shift uncertainty could have been reduced by operating at $\theta = 0$, the extremum of the quadrupole shift where sensitivity to the angular misalignment is reduced.  Other systematics are estimated to be $<1$ mHz and detailed in the following section. 
 
We find the quadratic Zeeman coefficients
\begin{align*}
\alpha_1 &= 2\,038.589\,1(31)~\mathrm{Hz}\,\mathrm{mT}^{-2}~\mathrm{and,}\\
\alpha_2 &= 2\,428.947\,0(26)~\mathrm{Hz}\,\mathrm{mT}^{-2}\mathrm{.}
\end{align*}   
The uncertainties are reduced by $\sim 14$ times compared to the previously reported values~\cite{zhiqiang2024}, and both are consistent to within $<$1.3$\sigma$ of the previous larger uncertainties.

\section{Quadrupole Shift Evaluation and Microwave Frequency Measurement\label{sec:results}}

\begin{figure}[t]
\begin{center}
\includegraphics[width=1.0\linewidth]{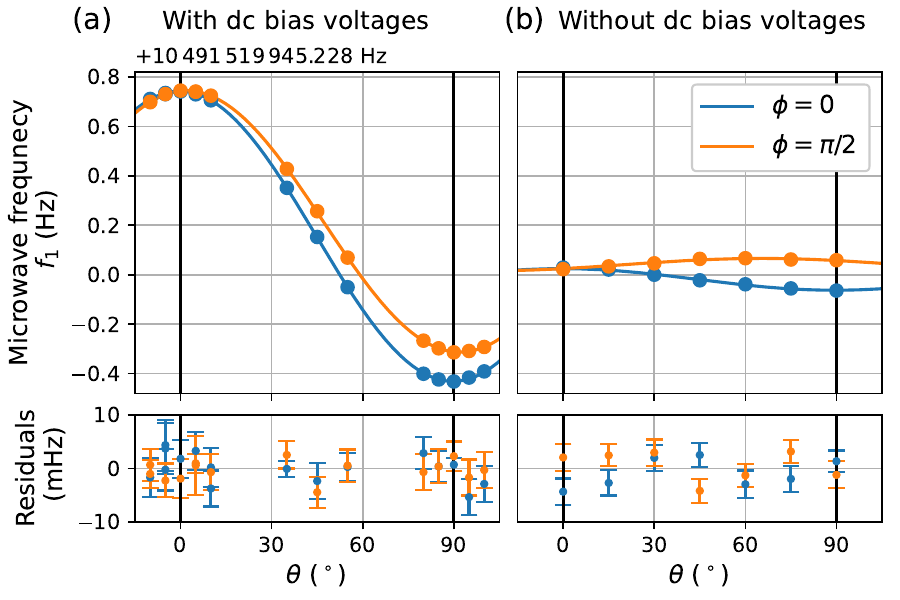}
\caption{\label{fig:quadrupole} Measurements of the ${^{3}}D_1$ $\ket{7,0}$ to $\ket{8,0}$ transition for a range of magnetic field orientations $(\theta,\phi)$ for (a) the typical applied dc biases, and (b) no applied dc bias. Solid lines are the fit to the quadrupole shift model given \eref{eq:quad}.  The reduced $\chi^2$ is $0.93$ with 25 degrees of freedom and 1.5 with 8 degrees of freedom for (a) and (b) respectively. Fit residuals are shown in the bottom plots.}
\end{center}
\end{figure}

We first determine the orientation of the electric field gradient by measuring the microwave frequency $f_1$ with the applied magnetic field oriented over the range of  $(\theta,\phi)$ accessible by our alignment procedure. All measurements are carried out at a magnetic field near 0.1 mT and with Ramsey interrogation time of $5 - 10\,$s. Results are shown in \fref{fig:quadrupole} for both the typical applied dc bias voltages and for no dc bias voltages.  Approximately 0.5 h of data collection was taken for each data point in order to average down to a statistical uncertainty of $\sim 2$ mHz. The quadratic Zeeman shift is evaluated from the magnetic field inferred from interleaved measurements and has been subtracted from the data. Other systematic shifts are $<1$ mHz. 

In order to model the data by \eref{eq:quad}, we allow for an arbitrary rotation $R(\theta_x,\theta_y,\theta_z) = X_{\theta_x} Y_{\theta_y} Z_{\theta_z}$ which relates the fixed-space frame to the principal-axis frame, in effect mapping the coordinates $(\theta,\phi)$ to $(\beta,\alpha)$. From numerical simulation of the ion trap with the typical dc voltages applied, we would expect $A\approx -53\,\mathrm{V}/\mathrm{cm}^{2}$, $\epsilon \approx 0.7$, and $(\theta_x,\theta_y,\theta_z) \approx (0,0,\frac{\pi}{4})$, which is to say the principal axes are aligned to the trap geometry. The fit for this case is shown in \fref{fig:quadrupole}(a) and the fit parameters $[\theta_x,\theta_y,\theta_z]$  $=$    $[1.7(1)^\circ,1.1(1)^\circ,39(2)^\circ]$, $A = -53.77(5)\,\mathrm{V}/\mathrm{cm}^{2}$, $\epsilon=0.71(0.20)$ are close to the predicated values with deviations attributed to stray gradients~\cite{tamm2009stray,dube2005electric}. 

In order to determine the unperturbed microwave frequencies with the highest possible accuracy, we reduce the amplitude of the quadrupole shift by removing the applied dc gradients. No dc offset bias are applied to electrode pairs, with only differential dc voltages applied as necessary to compensate micromotion.  The rf drive is also increased such that we obtain trap frequencies $2\pi \times (1190,1100,95)\,$ kHz from only the rf field confinement and any remaining stray dc gradients. As shown by the fit in \fref{fig:quadrupole}(b), this reduces the amplitude of the quadrupole shift by one order of magnitude, and the principal axes are no longer well aligned to the trap geometry.  

We determine the unperturbed absolute frequencies $f_1$ and $f_2$ by taking the average of measurements in the three orthogonal directions $(\hat{x},\hat{y},\hat{z})$ for the configuration with no applied bias voltages. For each orientation, $f_1$ and $f_2$ are measured in an interleave for approximately 5 hours each in order to average down to a statistical uncertainty of  $0.5\,$mHz per measurement. We find the unperturbed microwave transition frequencies to be

\begin{align*}
f_1 &= 10\,491\,519\,945.228\, {82\pm(32)_\mathrm{stat}\pm(23)_\mathrm{sys}}\,\mathrm{Hz},~\mathrm{and}\\
f_2 &= 11\,290\,004\,289.881\, {61\pm(32)_\mathrm{stat}\pm(20)_\mathrm{sys}}\,\mathrm{Hz}.
\end{align*}

The systematic effects and uncertainties are summarized in \tref{tab:sys}.  The systematics are evaluated to be less than the statistical uncertainty.  The leading systematics are excess micromotion (EMM) and incomplete cancellation of quadrupole shift due to the systematic uncertainty in the orthogonality of the three magnetic field orientations. The orientation of the principal axis frame as determined by the quadruple model fit shown in \fref{fig:quadrupole}(b) is used to evaluate the sensitivity to systematic errors in axes orientations. The quadrupole averaging uncertainty given in \tref{tab:sys} is estimated by Monte Carlo sampling allowing for 0.5$^\circ$ systematic uncertainty in the orthogonality of the $B$ field settings.

\begin{table}
    \footnotesize
    \caption{\label{tab:sys} Systematic corrections and uncertainties for the absolute frequency measurement of $^{176}\mathrm{Lu}^{+}$  ${^{3}}D_1$ microwave clock transitions $F = 7$ to $8$ ($f_1$) and $F = 7$ to $6$  ($f_2$). All offsets and uncertainties are given fractionally at $10^{-14}$ relative to the respective frequencies. $\dagger$ Quadratic Zeeman shifts are corrected point by point from interleaved field measurements.}

    \centering
    \begin{tabular}{lrrrr}
    \toprule
    &\multicolumn{2}{c}{$f_1$} & \multicolumn{2}{c}{$f_2$} \\
     \cmidrule(lr){2-3} \cmidrule(lr){4-5} 
 Systematic\quad\quad\quad\quad\quad\quad\quad\quad   &~~Corr.&\quad~~Unc.& ~~Corr.&\quad~~Unc. \\ 
    \midrule
Quadrupole averaging	&0.0 & 1.3 &0.0 & 1.1 \\ 
Excess micromotion (EMM)	&0.0 &1.7 &0.0 &1.4 \\ 
Maser accuracy	&-1.3 &0.5 &-1.3 & 0.5 \\ 
Microwave ac Zeeman	&-0.1 &0.1 &-0.1 &$0.2$ \\ 
Quadratic Zeeman             & $\dagger\,\,$ & 0.3 & $\dagger\,\,$ & 0.2\\
$B$ field evaluation	&0.9 &$<0.1$ &1.0 &$<0.1$ \\ 
rf ac Zeeman		&1.3 &0.1 &1.2 &0.1 \\ 
\midrule Total	&0.8 &2.2 &0.8 &1.9 \\ 
\bottomrule
   \end{tabular}
\end{table}

The contribution from excess micromotion (EMM) relevant to the microwave transitions is the tensor component of the ac Stark shift. The shift on the microwave transition frequency, $\Delta f_{i}$, may be written~\cite{berkeland1998minimization}
\begin{equation}
\Delta f_i = -\frac{1}{4h}   \Delta\alpha^{(2)}_i \left(\frac{m \Omega_\mathrm{rf}}{e} \right)^2 \langle v^2 \rangle (3 \cos^2{\zeta} -1),\label{eq:emm}
\end{equation}
where $v$ is the velocity of the ion due to EMM, $\zeta$ is the angle between the applied magnetic field and the direction of rf field at the ion, and $ \Delta\alpha^{(2)}_i$ is the differential tensor polarizability. The differential tensor polarizability on the $f_1$ transition, for example, is given by $  \Delta\alpha^{(2)}_1 = (C_{7,0} - C_{8,0}) \alpha^{(2)}$, where $\alpha^{(2)} = -4.40(34)$, in atomic units, is the dc tensor polarizability of the $^3D_1$ term~\cite{arnold2018blackbody} and the coefficients are as given by \eref{eq:cfmf}.  EMM is measured in three orthogonal directions using phase modulated sideband spectroscopy on the $^1S_0$ to $^3D_2$ optical clock transition~\cite{EMM}.  EMM was compensated before and after each microwave measurement in the three orthogonal directions $(\hat{x},\hat{y},\hat{z})$, to the level of $|\frac{\Delta f_i}{f_i}|\lesssim \times 1\times10^{-16}$ evaluated for the extreme ($\zeta=0$) field orientation. Just as for the quadrupole shift, the EMM tensor shift averaged over any three orthogonal orientations of the magnetic field is zero. However, variations in the amplitude or direction of the EMM during the microwave measurement campaign may result in a non-zero average. As the EMM compensation was not continuously monitored during the microwave spectroscopy, we take the full magnitude of the increase in EMM after each measurement as the systematic uncertainty of the EMM shift on each of the microwave frequencies.

The microwave synthesizers, based on the chip ADF5356, have $\sim 1~\mu $Hz resolution in the frequency setting and are referenced to a hydrogen maser. To assess the maser accuracy, the maser was evaluated with respect to the known Lu$^+$ ($^3D_1$) optical frequency standard~\cite{zhang2025absolute} before and after microwave frequency measurements. The instability of the maser has been previously characterized~\cite{zhang2025absolute} and individual measurements against Lu$^+$ ($^3D_1$) are limited by the $\sim4\times 10^{-15}$ noise floor of the maser for typical integration times.  The interpolated fractional frequency offset of the maser is evaluated to be  {$-1.3(0.5)\times10^{-14}$} over the duration of the microwave measurements. The maser offset was evaluated against UTC over the same interval using GNSS methods~\cite{jian2023gps} to be $-1.2(0.3)\times10^{-14}$, consistent with evaluation against Lu$^+$ ($^3D_1$).  The maser has $\sim2\times10^{-14}$/K sensitivity to the ambient temperature and resides in the same lab as the Lu$^+$ apparatus. The lab temperature is continuously monitored and had variation $<0.1$K relative to the mean temperature over the duration of the microwave measurements. Thermal-driven fluctuations in the maser are estimated to be well below its characterized instability. 

At this level of precision, cycle synchronous, or thermal, phase variations are a potential concern. The microwave synthesizer outputs are switched by external high-speed solid state switches and amplified before delivery to the microwave horns.  We checked for phase chirp as result of switching by comparing the phase~\cite{sherman2016oscillator} of the microwaves delivered to the horn to an independent, commonly referenced, synthesizer. Additionally, measurements of the $^3D_1$ microwave transition frequency were performed over a range of Ramsey times to check for phase chirp induced shifts, or other microwave probe induced shifts, which would to vary as $\sim1/T$ with Ramsey time. To highlight the importance of such measurements, we note that a shift was detected in this way initially for different synthesis configuration and subsequently able to be resolved.  For the synthesis configuration used in the reported experiments, we saw no evidence of significant phase chirp in either test.  We consider thermally-driven phase effects on the synthesis and distributed 10 MHz the maser reference signal. The synthesizer phase instability is bound to $\lesssim 1\times10^{-11} (\tau/\mathrm{s})^{-1}$ by comparison to an independent synthesizer. The reference distribution phase instability is estimated by accounting for the propagation delay in $\sim$20 m of coaxial cable (-0.8 $\mathrm{ps}\,\mathrm{m}^{-1}\,\mathrm{K}^{-1} $) and the $\lesssim0.1$ K ambient temperature stability.  Both effects contribute well below the measurement instability. 

The microwave fields used for Ramsey spectroscopy contribute an ac-Zeeman shift. The frequency shift is largely suppressed by the long (5 s) Ramsey time compared to the short ($\sim$5 ms) $\frac{\pi}{2}$ pulse time. For some measurements the shift is as large as 100 $\mu$Hz when the microwave polarization is particularly imbalanced. Each time the magnetic field orientation was changed, the microwave field polarization was reevaluated by comparing the coupling strength on $m_F=0$ to $m_F^\prime = [-1,0,1]$ microwave transitions. Expressions for evaluation of the probe-induced microwave ac-Zeeman shift are given in the Supplemental Material of~\cite{zhiqiang2024}.

The quadratic Zeeman coefficients $\alpha_i$ reported in \sref{sec:qzc}  contribute $\sim3\times10^{-15}$ fractional uncertainty at a 0.1 mT magnetic field. The magnetic field is inferred from the Zeeman splitting of the $\ket{6,\pm1}$ states as measured by Rabi spectroscopy on the $\ket{7,0}$ to $\ket{6,\pm1}$ microwave transitions in interleaved measurements. An imbalance in the $\sigma^\pm$ polarization components or a relatively large $\pi$ component of this microwave field gives rise to a differential ac Zeeman shift on these lines and consequently an error, $\delta B$, in the inferred magnetic field. Detailed calculations on the determination of $\delta B$ are given in Appendix \ref{fielderr}. The systematic shift on the microwave clock transition frequency is $\approx 2 \alpha_i B \delta B$ and evaluates to $\lesssim 100\,\mu$Hz. 

 Electric currents to the trap electrodes driven by the rf-trapping potential give rise to an rf-magnetic field and corresponding ac-Zeeman shift. The shift depends on the component of the rf-field perpendicular to the applied dc field~\cite{gan2018oscillating}. We determine both the amplitude and orientation of the rf-magnetic field by measuring the Autler-Townes splitting on the Ba$^+$ clock transition~\cite{arnold2020precision} for the applied magnetic field in each of the three orthogonal directions. We determine the rf magnetic field component amplitudes to be $$[B_x,B_y,B_z] = [0.461(14), 0.3081(93), 0.2216(66)]\,\mu\mathrm{T}.$$ The systematic shift on the microwave clock transition is $\sim 200\,\mu$Hz for all three measurement orientations.

The rf electric field gradients used to trap the ion also give rise to a second-order ac-quadrupole shift, which does not average to zero over three mutually orthogonal magnetic field orientations~\cite{arnold2019oscillating}. This shift is evaluated to be $<2\times10^{-16}$ fractionally and so is not included in \tref{tab:sys}.

\section{Hyperfine mediated effects}

\begin{figure}[t]
\begin{center}
\includegraphics[width=1.0\linewidth]{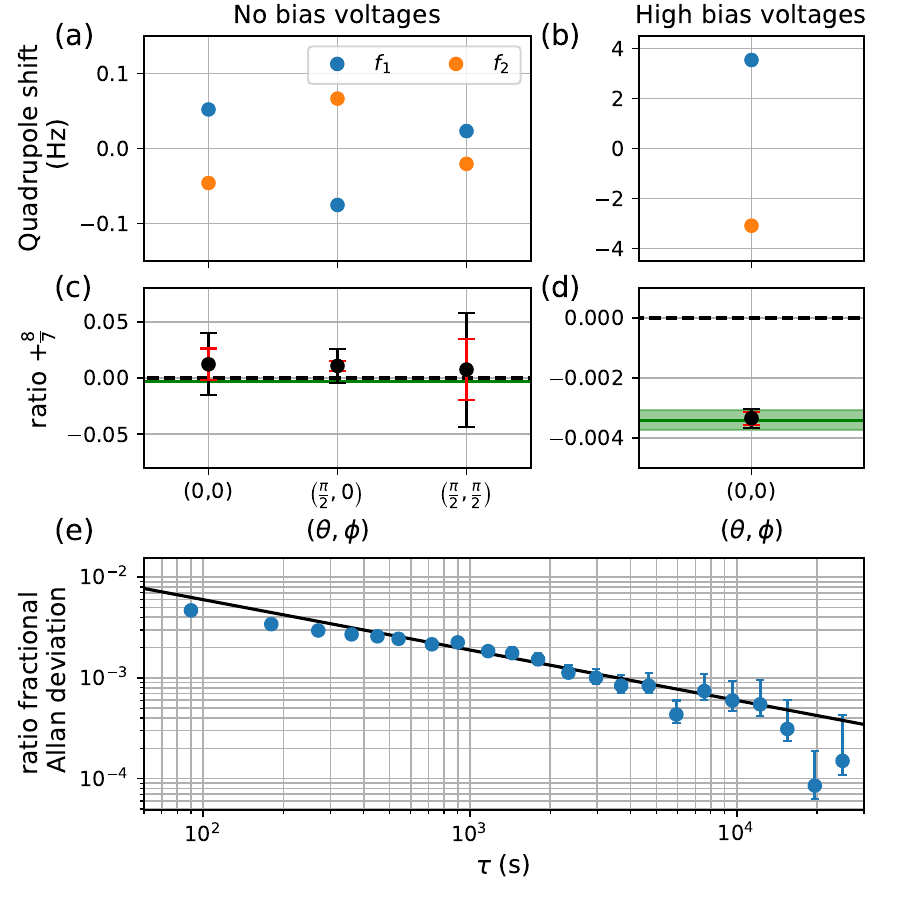}
\caption{\label{fig:results} (a,b) Quadrupole shifts measured with no bias applied (a) and with high bias applied (b). Error bars are smaller than the points. (c,d) measured ratio of quadruple shift compared to the expected value (green line) based on the evaluation hyperfine mediated effects in \cite{zhiqiang2020hyperfine} (\eref{eq:ratio}). (e) Allan deviation of ratio measurement at high bias.}
\end{center}
\end{figure}

As motivated in \sref{sec:theory}, we consider the ratio of quadrupole shifts on transitions $f_1$ and $f_2$. \fref{fig:results}(a,c) shows the quadrupole shifts and ratios in the three directions from the frequency measurement data where no bias voltages were applied. Here the quadrupole shifts are small, so the fractional uncertainty in ratio is large compared to the hyperfine mediated effects. The ratios are statistically consistent for all three orientations suggesting there is no orientation dependent systematic unaccounted for. To increase the precision on the ratio measurement, we apply a high bias voltage of 38 V on the endcap electrodes which gives an axial trapping frequency of  370 kHz. From  22 hours of measurement time with the magnetic field aligned to the trap axis, we find the quadrupole shifts and ratio shown in \fref{fig:results}(b,d). 

The deviation from $-\frac{8}{7}$ is clearly observed and from \eref{eq:ratio} we can directly infer 
\begin{equation}
\beta_{1,2}^{Q}= [- {0.01304}\pm(91)_\mathrm{stat} \pm (85)_\mathrm{sys} ] e a_0^2
\end{equation}
consistent with the value reported in ~\cite{zhiqiang2020hyperfine}. This gives an experimental determination of 
\begin{equation}
\langle \delta \Theta \rangle_{F} \approx \frac{2}{105} \beta_{1,2}^{Q} =   [ {-2.48}\pm(17)_\mathrm{stat} \pm (16)_\mathrm{sys} ] \times 10^{-4} e a_0^2.
\end{equation}
for the RQM in agreement with the value $-2.54(25)\times10^{-4}\,e a_0^2$ given in~\cite{zhiqiang2020hyperfine}, which was indirectly determined from $g$-factor measurements and theoretically calculated matrix elements.

\section{Discussion}

In summary we have measured both $^{176}\mathrm{Lu}^+$ $^3D_1$ microwave transition frequencies to fractional accuracy of $4\times10^{-14}$. Although the motivation for this work was not to develop a microwave frequency standard, for comparison we note that the fractional frequency accuracy is only a factor of two to four worse than the highest accuracies reported for frequency standards based on the microwave transitions in trapped ion systems, including Cd$^+$~\cite{qin2022high}, Hg$^+$~\cite{berkeland1998laser}, and Yb$^+$~\cite{phoonthong2014determination}. 

The unperturbed microwave frequencies measured here enable measurement of the $^3D_1$ microwave quadrupole shifts, providing a simple and accurate means to evaluation the quadrupole shift in $^{176}\mathrm{Lu}^+$ ($^3D_1$) optical frequency standards. 

We extended this application to measurement of the ratio of quadrupole shifts with sufficient precision to infer the RQM resulting from hyperfine mediated effects. The inferred RQM is in agreement with the previous evaluation~\cite{zhiqiang2020hyperfine}. Consistency in results from the two different approaches provide greater confidence in the RQM assessment for the $^{176}\mathrm{Lu}^+$ ($^3D_1$) optical frequency standard.

\begin{acknowledgments}
This project is supported by the National Research Foundation, Singapore through the National Quantum Office, hosted in A*STAR, under its National Quantum Engineering Programme 3.0 Funding Initiative (W25Q3D0007) and under its Centre for Quantum Technologies Funding Initiative (S24Q2d0009).

The data that support the findings of this article are openly available~\cite{datasets}.

\begin{appendix}

   \section{Field determination error}
\label{fielderr}
Given that the quadratic Zeeman shift is the largest shift on the microwave clock frequency, any errors in the determination of the magnetic field through the Zeeman servo can lead to shifts $\sim 100\,\mu$Hz. There is a bias in the magnetic field assessment depending on the polarization of the microwave field. We adopt the notation $a_{k,q}$ to refer to the $q$ polarization component of the horn driving the microwave transition $k$, where $\sum_q a_{k,q}^2 = 1$. The microwave polarizations are determined by ratios of pi-times on the $\pi,\sigma_\pm$ transitions.
\begin{equation}
\frac{a_{1\pm}}{a_{10}} = \frac{4}{3} \frac{\tau_{10}}{\tau_{1\pm}} \qquad \frac{a_{2\pm}}{a_{20}} = \sqrt{\frac{7}{3}} \frac{\tau_{20}}{\tau_{2\pm}}
\end{equation}
When probing the Zeeman pair $\ket{7,0}$ to $\ket{6,\pm 1}$ to determine the magnetic field, there is a probe-induced shift due to off-resonant coupling primarily from $\ket{6,\pm 1}$ to $\ket{7,\pm 1}$, but also $\ket{6,\pm 1}$ to $\ket{7,\pm 2}$ and $\ket{7,0}$ to $\ket{6,0}$. The shifts on each level are given as follows
\begin{align}
\Delta_{7,0}^{(\pm)} &= \mp \frac{(\Omega_{7060})^2}{4\omega_6} \mp \frac{(\Omega_{706(\mp1)})^2} {4(2\omega_6)} \nonumber \\ &= \mp \frac{(\Omega_{7060})^2}{4\omega_6} \left( 1 + \frac{3}{14}\left(\frac{a_{2\mp}}{a_{20}} \right)^2 \right) \label{eqn:shift_70}\\
\Delta_{6,\pm1}^{(\pm)} &= \mp \frac{(\Omega_{7(\pm1) 6(\pm1)})^2}{4\omega_7} \mp \frac{(\Omega_{7(\pm2) 6(\pm1)})^2}{4(2\omega_7)} \nonumber \\ &= \mp \frac{(\Omega_{7060})^2}{4\omega_7} \left( \frac{48}{49} + \frac{18}{49}\left(\frac{a_{2\mp}}{a_{20}} \right)^2 \right) \label{eqn:shift_61}
\end{align}
where $\hbar\omega_F = |g_F \mu_B B|$ are the Zeeman splittings and $\Omega_{F m F^\prime m^\prime}$ is the coupling from $\ket{F,m}$ to $\ket{F^\prime,m^\prime}$. The overall shift when probing the $\ket{7,0}$ to $\ket{6,\pm 1}$ transition is given by $\Delta_{6,1_\pm}^{(\pm)} - \Delta_{7,0}^{(\pm)}$. The shift in the linear Zeeman splitting and the magnetic field are given as follows
$$ \delta B =  \frac{\hbar}{\mu_B g_6} \Delta \omega_6 = \hbar \frac{(\Delta_{6,+1}^{(+)} - \Delta_{7,0}^{(+)}) - (\Delta_{6,1_-}^{(-)} - \Delta_{7,0}^{(-)})}{2\mu_B g_6}.$$
The systematic shift on the microwave clock transition frequency is
$$ \Delta f_k = \alpha_k (B+\delta B)^2 - \alpha_k B^2  \approx 2\alpha_k B \delta B. $$
Similarly we have the following expressions when probing the $\ket{7,0}$ to $\ket{8,\pm 1}$ Zeeman pair to determine the magnetic field:
\begin{align}
\Delta_{7,0}^{(\pm)} &= \mp \frac{(\Omega_{7080})^2}{4\omega_8} \mp \frac{(\Omega_{708(\mp1)})^2}{4(2\omega_8)} \nonumber \\ &= \mp \frac{(\Omega_{7080})^2}{4\omega_8} \left( 1 + \frac{9}{32}\left(\frac{a_{1\mp}}{a_{10}} \right)^2 \right) \label{eqn:shift_7080}\\
\Delta_{8,\pm1}^{(\pm)} &= \mp \frac{(\Omega_{7(\pm1) 8(\pm1)})^2}{4\omega_7} \mp \frac{(\Omega_{7(\pm2) 8(\pm1)})^2}{4(2\omega_7)} \nonumber \\&= \mp \frac{(\Omega_{7080})^2}{4\omega_7} \left( \frac{63}{64} + \frac{21}{128}\left(\frac{a_{1\mp}}{a_{10}} \right)^2 \right) \label{eqn:shift_817m}.
\end{align}
For this case, the shift in the linear Zeeman splitting is
$$ \delta B =  \frac{\hbar}{\mu_B g_8} \Delta \omega_8 = \hbar \frac{(\Delta_{8,+1}^{(+)} - \Delta_{8,0}^{(+)}) - (\Delta_{8,-1}^{(-)} - \Delta_{8,0}^{(-)})}{2\mu_B g_8}.$$

\end{appendix}

\end{acknowledgments}
%

\end{document}